\documentclass[prl,aps,twocolumn,showpacs]{revtex4}

\usepackage{subfigure}
\usepackage{color,amsmath,graphicx,epsfig,amssymb}
\usepackage[T1]{fontenc}
\usepackage{pslatex}
\usepackage{txfonts}

\begin{document}

\def\pd#1#2{\frac{\partial #1}{\partial #2}}
\def\bk{{\bf k}}
\def\bx{{\bf x}}
\def\grad{{\mbox{\boldmath$\nabla$\unboldmath}}}

\title{\bf Breathers on Quantized Superfluid Vortices}
\author {Hayder Salman}
\affiliation {School of Mathematics, University of East Anglia, Norwich Research
Park, Norwich, NR4 7TJ, UK}

\begin {abstract}
We consider  the propagation of breathers along a quantised superfluid vortex.
Using the correspondence between the local induction approximation (LIA) and the
nonlinear Schr\"odinger equation, we identify a set of initial conditions
corresponding to breather solutions of vortex motion governed by the LIA. These initial conditions, which
give rise to a long-wavelength modulational instability, result in the emergence
of large amplitude perturbations that are localised in both space and time. The
emergent structures on the vortex filament are analogous to loop solitons. 
Although the breather solutions we study are exact solutions of the
LIA equations, we demonstrate through full numerical simulations that their key
emergent attributes carry over to vortex dynamics governed by the Biot-Savart
law and to quantized vortices described by the Gross-Pitaevskii equation.
The breather excitations can lead to self-reconnections, a mechanism that can play an important role within the cross-over range of scales in superfluid
turbulence. Moreover, the observation of breather solutions on vortices in a field model suggests that these solutions are expected to arise in a wide range of other physical contexts from classical vortices to cosmological
strings.
\end{abstract}
\maketitle

The prediction of nonlinear wave phenomena such as solitons and breathers in a
variety of different contexts has spurred an active search for  these
types of waves. Solitons, which are solutions of integrable equations, with
an infinite number of constants of motion, are  of interest since they are localised
disturbances transferring energy over large distances. Moreover, solitons retain
their shape upon interacting with other solitons thereby ensuring their tendency
to persist once formed. It has been shown that solitons are
paradigms for studying Tsunamis \cite{Slunyaev11} in the oceans. In superfluids, dark solitons can give rise
to quantised vortices through the so-called snake instability \cite{Brand02}, and the
interaction of bright solitons has been extensively studied particularly in
nonlinear optics \cite{Ablowitz11}. Although soliton solutions  do not always exist in
higher dimensions, solitary wave disturbances, for example in the form of vortex
dipoles or vortex rings in superfluids \cite{Jones82}, can often be found which retain some of the
stability properties of their 1D counterparts.

In contrast to solitons, breather solutions of
integrable partial differential equations are less studied despite their generic properties in a wide range of nonlinear systems. Breather solutions were first predicted by Kuznetsov \cite{Kuznetsov77} and Ma \cite{Ma79}, Akhmediev \cite{Akhmediev87}, and Peregrine \cite{Peregrine83} as localised disturbances in both space and time. However, they were only observed very recently in \cite{Kibler10}.
Very soon afterwards, further confirmation came from studies of surface waves \cite{Chabchoub11} and
the observation of the Kuznetsov breather was later reported in \cite{Kibler12}. 
In water waves, breathers have been used as paradigms of freak or rogue waves in the oceans 
\cite{Slunyaev11,Onorato11}. 

There exits a very interesting connection between breathers and vortex dynamics that has not been fully explored. It is known that the motion of a vortex filament in the local induction approximation (LIA) can be mapped through the Hasimoto transformation \cite{Hasimoto72} onto the nonlinear Schr\"{o}dinger equation of the self-focussing type. 
This connection was exploited in \cite{Hopfinger82} to produce solitary waves propagating along highly concentrated vortex cores within a classical turbulent flow. The success of the Hasimoto theory in explaining these observations motivated further work in \cite{Cieslinski92,Levi83,Fukumoto86,Konno91,Konno92} to study the motion of such solitons on vortices. The Hasimoto loop soliton on a vortex filament was also found to exhibit a striking resemblance to loop like excitations observed on the funnel of a tornado (see photo in \cite{Aref84}). However, given that the loop emerges in a time-dependent manner, the vortex loop on the tornado may actually be an example of a breather rather than a
Hasimoto soliton. 

Loop-like excitations on quantized vortices are relevant to the study of vortices in superfluids.
In fact, uncovering new mechanisms of vortex dynamics is likely to
contribute to our understanding of superfluid turbulence, a field which has been
vigorously investigated  but one that continues to pose major challenges in our understanding
at a fundamental and conceptual level \cite{Vinen08,Tsubota09}. We will, therefore, model the breather solutions on quantised vortices using a hierarchy of different
models.

We begin by recalling that a quantized vortex with circulation $\Gamma$ induces
a velocity according to the Biot-Savart law given by
\begin{eqnarray}
\mathbf{v}(\mathbf{R}) = \frac{\Gamma}{4\pi} \int_{\mathcal{C}(t)}
\frac{(\mathbf{R}-\mathbf{r}) \times d\mathbf{r}}{|\mathbf{R}-\mathbf{r}|^3},
\end{eqnarray}
where $\mathbf{R}$ is the position vector for any point in the fluid, and the vector $\mathbf{r}$ runs along the vortex filament parameterised by the arclength $s$. The integral is evaluated along the curve $\mathcal{C}(t)$ that varies with time. This equation shows that the velocity induced on a superfluid vortex depends on
the instantaneous configuration of a vortex line where non-local contributions arise from the terms under the integral. To simplify the problem, it can be shown \cite{Donnelly91} that, at leading order, the vortex velocity depends on the local radius of curvature $R$ and the vortex core radius $a_o$ with
\begin{eqnarray}
d\mathbf{r}/dt = (\Gamma/4\pi) \ln (R/a_o) \mathbf{r}'\times
\mathbf{r}'' \equiv \beta (\mathbf{r}' \times \mathbf{r}'') = \beta \kappa \mathbf{b},
\label{eqn_LIA}
\end{eqnarray} 
where primes denote differentiation with respect to arclength.
By making use of the Seret-Frenet equations given by \cite{Pismen99}, 
\begin{eqnarray}
\mathbf{r}' = \mathbf{t}, \;\;\; \mathbf{t}' = \kappa \mathbf{n}, \;\;\;
\mathbf{n}' = \tau \mathbf{b} - \kappa \mathbf{t}, \;\;\; \mathbf{b}' = -\tau
\mathbf{n}, \label{eqn_SF}
\end{eqnarray}
where $\mathbf{t}, \mathbf{n}, \mathbf{b}$ are a right-handed system of mutually
perpendicular unit vectors corresponding to the tangent, the principal normal
and binormal directions, respectively, Eq.\ (\ref{eqn_LIA}) can be written as
$\dot{\mathbf{r}} = \kappa \mathbf{b}$. This local induction approximation (LIA)
allows us to obtain much more insight into the properties of solutions associated with the motion of a superfluid vortex. 
The most significant insight that has emerged from using the LIA was
obtained in \cite{Hasimoto72}. By assuming $\beta$ is constant,  it was shown that Eq.\ (\ref{eqn_LIA}) can be mapped onto a 1D nonlinear Schr\"odinger equation of a self-focusing type given by
\begin{eqnarray}
\beta^{-1} (i\phi_t) = -\phi_{ss} - \frac{1}{2}|\phi|^2 \phi, \label{eqn_1DNLS}
\end{eqnarray}
where the complex function $\phi(s,t)$ is a function of arclength $s$ and time
$t$, and is related to the intrinsic parameters of the vortex given by the local
curvature $\kappa(s)$ and the torsion $\tau(s)$ through the transformation
$\phi(s,t)= \kappa \exp (i\int \tau(s') ds')$. Since
the 1D NLS equation is integrable \cite{Ablowitz11}, its soliton solutions correspond to 
large amplitude excitations propagating along a vortex filament. The interactions of these solitons on a
vortex are of interest since they can promote a self-reconnection and drive the
emission of vortex rings from the vortex as suggested in \cite{Svist95} in the scenario 
of superfluid turbulence. However, Eq.\ (\ref{eqn_1DNLS}) has an even more
interesting set of solutions corresponding to breathers since it is an NLS
equation of the self-focusing type. Unlike solitons, breathers are
localised disturbances that appear unsteady in any frame of reference. Breathers
typically come in different types. In this work, we will be primarily interested in vortices that have a
periodic structure along the arclength. For this reason, we will focus on the
Akhmediev breather solution, which will turn out to be more relevant for our
considerations.

The Akhmediev breather corresponds to, so-called, heteroclinic solutions of the NLS equation \cite{Akhmediev86} although some authors refer to them as homoclinic solutions \cite{Ablowitz90} despite the presence of a phase-shift between the initial and final states. These nonlinear solutions are now
understood to be associated with the nonlinear stages of the Benjamin-Feir (or side-band) instability \cite{Benjamin67a,Benjamin67b,Samuels90}, which was also independently discovered in \cite{Zakharov68}.
The Akhmediev breather solution is typically cast in a form that corresponds to an orbit that is homoclinic/heteroclinic to a spatially uniform plane wave fixed point. In terms of a vortex, the basic state of this solution has no torsion and so would not correspond to vortices that are of general interest in superfluids. What is  needed is a solution that is
homoclinic/heteroclinic to a travelling plane wave solution which would correspond to a helical vortex. 
The required solutions can be derived directly from the breather
solutions that are homoclinic/heteroclinic to the spatially uniform plane wave solution using an
invariance property of the underlying NLS equation. We note that Eq.\ (\ref{eqn_1DNLS})
is invariant under the Galilean (gauge) and scaling transformations given respectively by
\begin{eqnarray*}
&& t \rightarrow t, \;\;\;\; s \rightarrow s-2\tau_o t, \;\;\;\; \phi \rightarrow \phi \exp [i(\tau_o s - \tau_o^2 t)] \\
&& t \rightarrow \kappa_o^2t, \;\;\;\; s \rightarrow \kappa_o s, \;\;\;\; \phi \rightarrow \kappa_o \phi (\kappa_o s, \kappa_o^2 t),
\end{eqnarray*}
A spatially uniform plane wave solution of Eq.\ (\ref{eqn_1DNLS}) given by $\phi = \kappa_o
\exp [i(\kappa_o^2t/2]$, and corresponding to a vortex ring, then transforms to the solution describing a helical vortex. This is given by $\phi = \kappa_o \exp [i\tau_o s+i(\kappa_o^2-\tau_o^2)t]$ where $\kappa_o$ are $\tau_o$ are constants denoting the curvature and torsion of the unperturbed helical vortex filament, respectively. It follows that the
breather solution we seek is given by \cite{Dysthe99}
\begin{eqnarray}
&&\phi_A(s,t) = \kappa_o \mathrm{e}^{i\kappa_o^2 \beta t/2+i\tau_o s+i(\kappa_o^2-\tau_o^2)\beta t} \times \nonumber \\
&& \;\;\;\;\;\;\;\;\;\;\;\frac{\cosh (\Omega \kappa_o^2 \beta t/4 - 2i\varphi)-\cos (\varphi)
\cos (k \kappa_o s/2-k \tau_o \kappa_o \beta t)}{\cosh (\Omega \kappa_o^2 \beta t/4) - \cos(\varphi)\cos(k \kappa_o s/2-k \tau_o \kappa_o \beta t)}, \nonumber \\
 && k = 2\sin(\varphi), \;\;\;\;\;\; \Omega = 2\sin(2\varphi). \label{eqn_Akh}
\end{eqnarray}
To recover the instantaneous shape of the
filament, we integrate the Serret-Frenet equations subject to the initial conditions evaluated at some point
$\mathbf{t}(s_p(t))$, $\mathbf{n}(s_p(t))$, $\mathbf{b}(s_p(t))$. The point $s_p$ is chosen to coincide with a point of maximum curvature such that $d\kappa(s)/ds = 0$.  Since the breather solution corresponds to a wave travelling along the filament with speed $2\tau_o$, we set $s_p(t) = 2\tau_o(t-t_i) + s_p(t_i)$ with $t_i$ denoting an initial time where the breathing mode is asymptotically small.

While this specifies the filament up to an arbitrary orientation, we must also calculate the orientation of the tangent, normal, and binormal vectors at the current time $t$, given that the orientation of the filament is specified at the initial time $t_i$. We accomplish this by casting the equations describing the rate of change of these vectors with time at a fixed point $s$ in the form described by Hasimoto \cite{Hasimoto72}. By introducing the variables $\mathbf{N} = (\mathbf{n}+i\mathbf{b}) \exp(i \int \tau_{s_p}^s d\tilde{s})$ and $\psi = \kappa \exp(i \int \tau_{s_p}^s d\tilde{s})$, the respective equations of motion are given by
\begin{eqnarray}
\dot{\mathbf{t}}  = \frac{1}{2} \left( \psi' \overline{\mathbf{N}} - \overline{\psi'} \mathbf{N} \right), \;\;\;\;\;
\dot{\mathbf{N}}  = \frac{i}{2} \left( \kappa^2 \mathbf{N} - 2\psi' \mathbf{t} \right), \label{eqn_tn}
\end{eqnarray}
where bars denote complex conjugate quantities and $\int d\tilde{s}$ denotes integration along arclength.

Since we are interested in the rate of change of these quantities along the point $s_p(t)$, our equations of motion become
\begin{eqnarray}
\frac{D}{Dt} \left( \begin{array}{c} 
\sqrt{2} \mathbf{t} \\ \mathbf{N} \\ \overline{\mathbf{N}}
\end{array} \right) = \left(  \begin{array}{ccc}
\mathbf{0} & -\overline{h} \mathbf{I} & -h \mathbf{I} \\
 h \mathbf{I} & \frac{i\kappa^2}{2} \mathbf{I} & \mathbf{0} \\
\overline{h} \mathbf{I} & \mathbf{0} & -\frac{i\kappa^2}{2} \mathbf{I} 
\end{array}
\right) \left( \begin{array}{c} 
\sqrt{2}  \mathbf{t} \\ \mathbf{N} \\ \overline{\mathbf{N}}
\end{array} \right), \label{eqn_SF}
\end{eqnarray}
where, $h(s) = -\left( i \psi'/\sqrt{2} + \sqrt{2} \tau_o \psi \right)$, $\frac{D}{Dt} = \left. \frac{\partial}{\partial t} \right|_s + 2\tau_o \left. \frac{\partial}{\partial s} \right|_t$, and $\mathbf{0}$ and $\mathbf{I}$ are $3 \times 3$ zero and identity matrices, respectively. We have written the governing equations in a form that emphasises their skew-Hermitian structure which arises as a consequence of the underlying symplectic form of the LIA. 

\begin{figure}[t]
\centering
 \begin{minipage}[b]{0.49\textwidth}
    \centering
    \subfigure[\label{fig_t0} \hspace{0.3cm} t = -800 \hspace{1.75cm} t=-20 \hspace{1.85cm} t=-0.1 \hspace{0.8cm}]{
      \label{fig:mini:subfig:a}
      \includegraphics[width=0.95in]{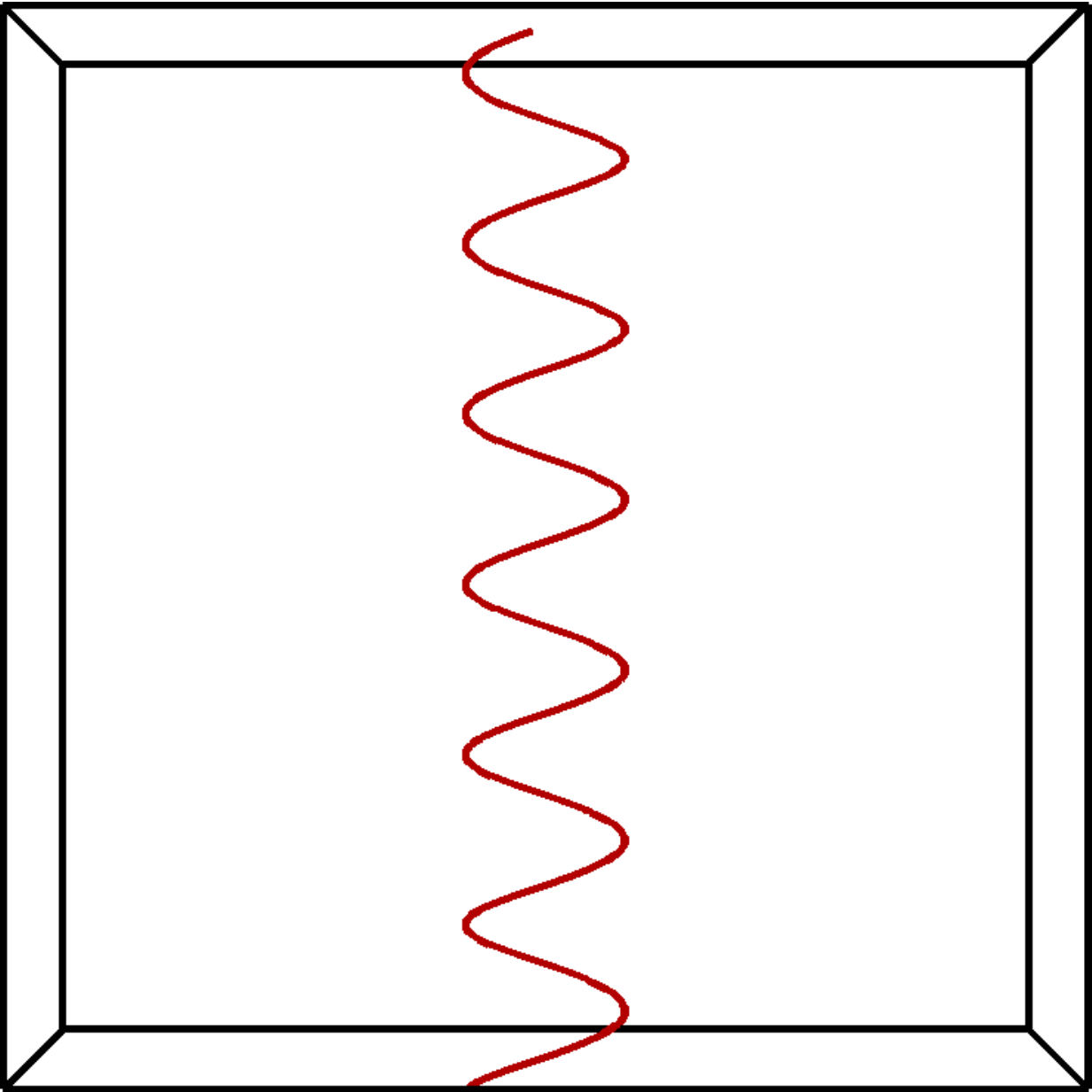}
      \hspace{0.2cm}
      \includegraphics[width=0.95in]{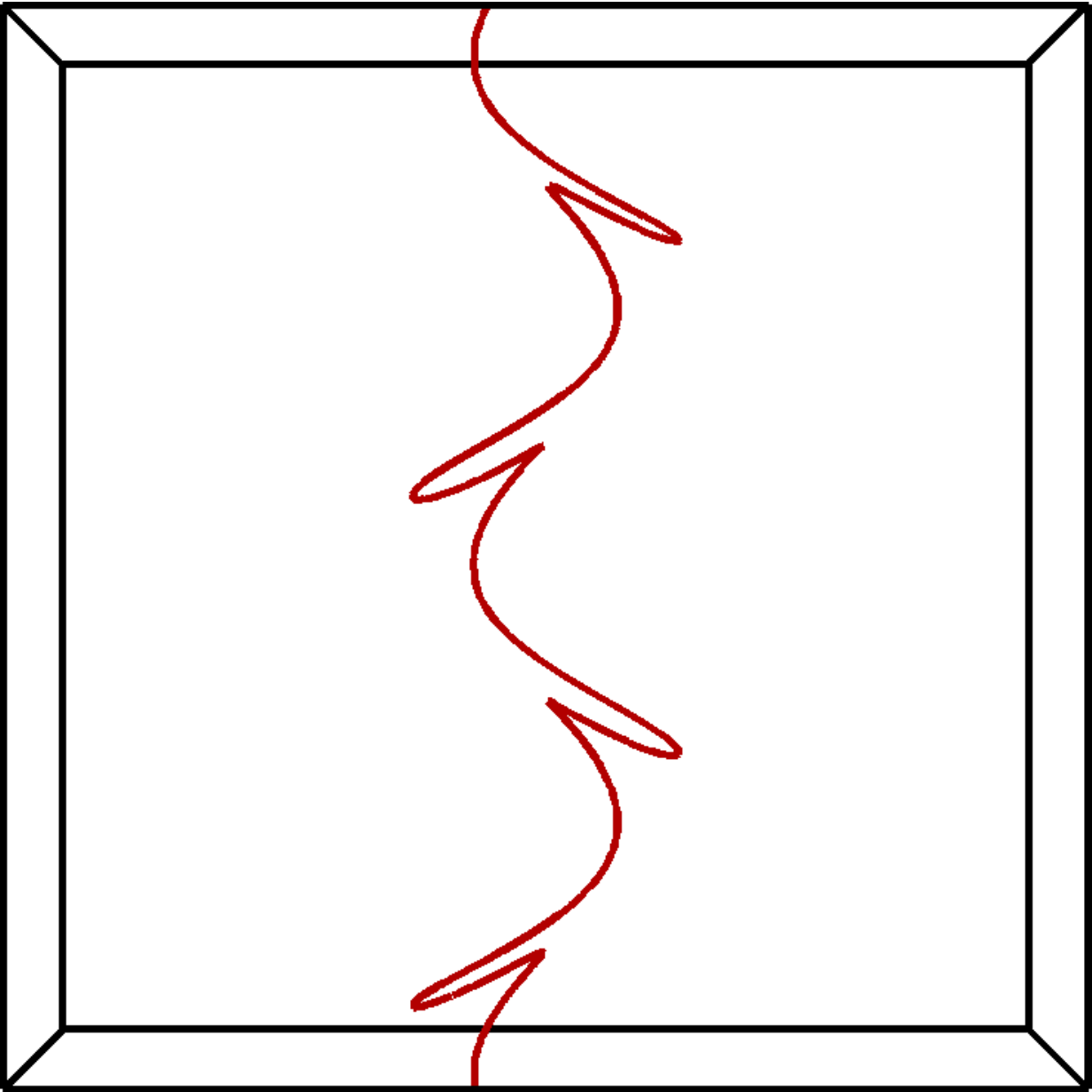}
      \hspace{0.2cm}
      \includegraphics[width=0.95in]{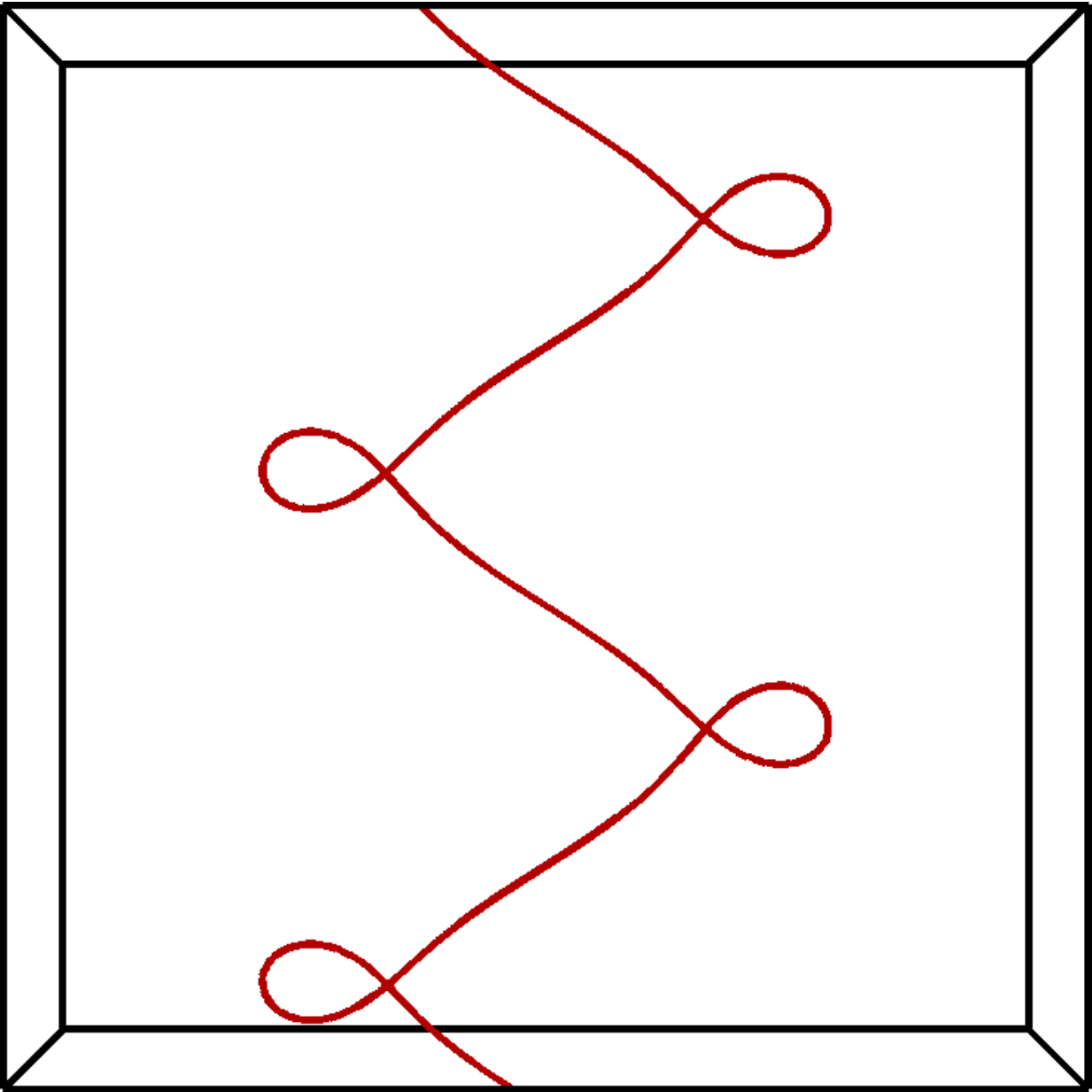}}
  \end{minipage}
 \begin{minipage}[b]{0.49\textwidth}
    \centering
    \subfigure[\label{fig_t0} \hspace{0.3cm}  t = 100 \hspace{1.75cm} t=160 \hspace{1.85cm} t = 240 \hspace{0.8cm}]{
      \label{fig:mini:subfig:a}
      \includegraphics[width=0.95in]{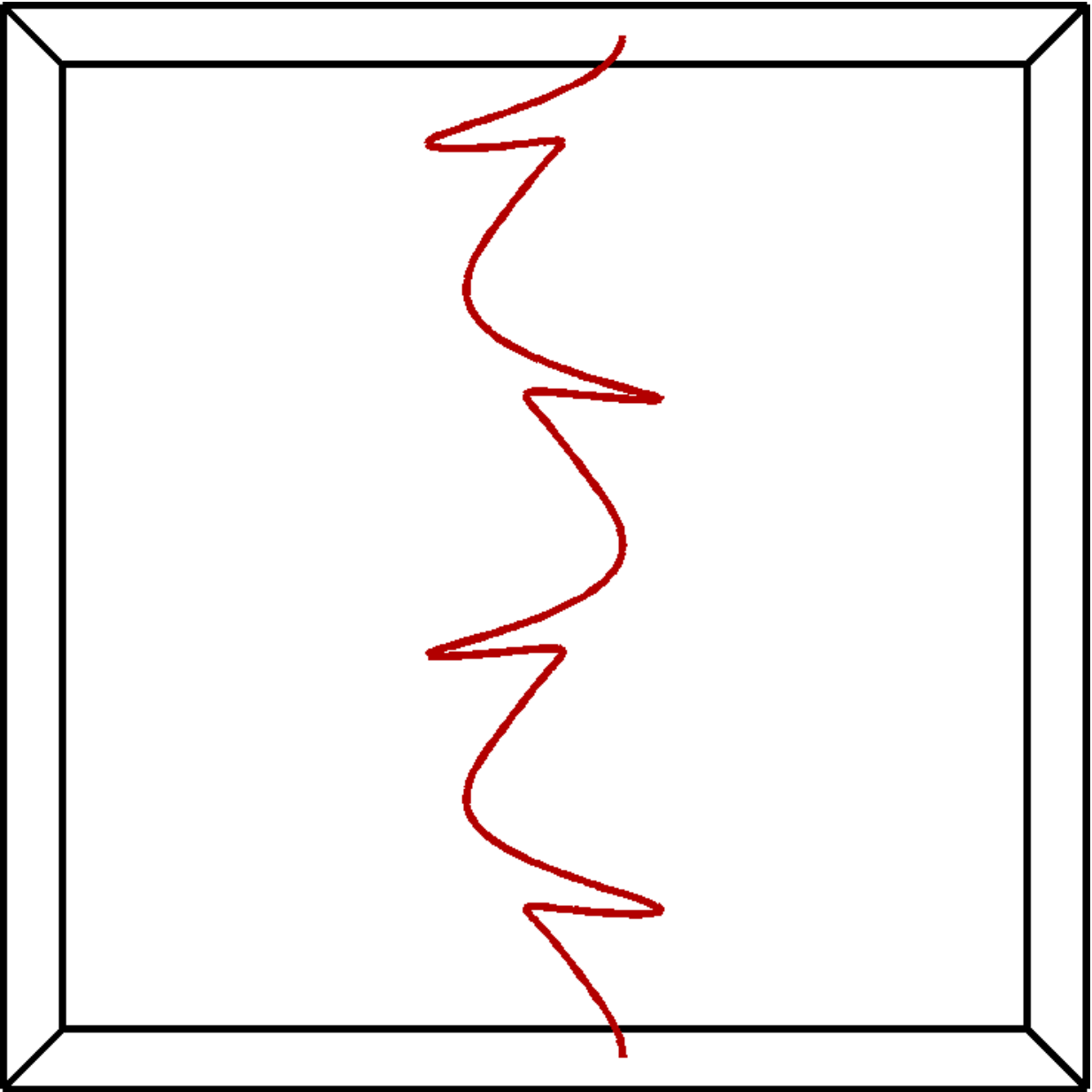}
      \hspace{0.2cm}
      \includegraphics[width=0.95in]{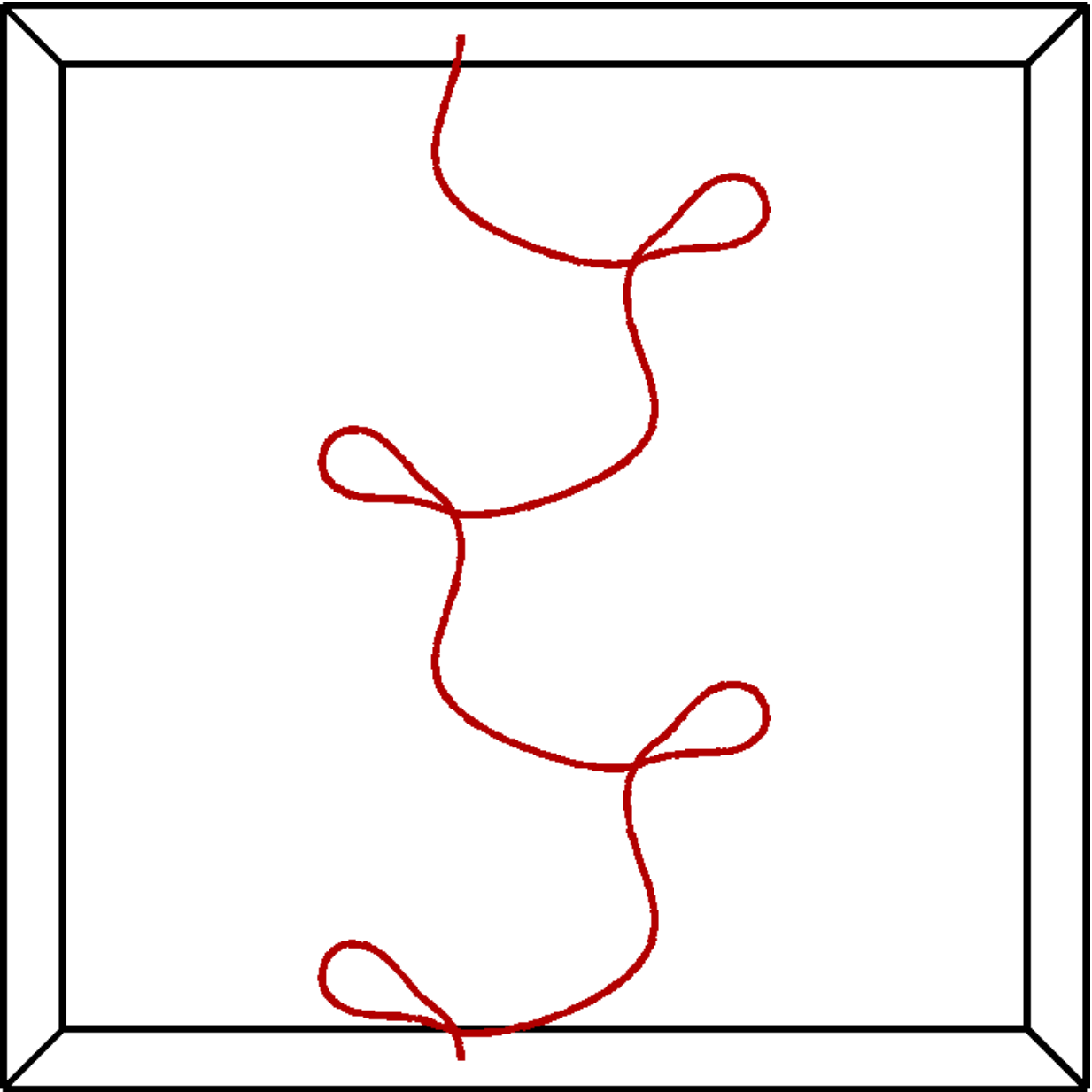}
      \hspace{0.2cm}
      \includegraphics[width=0.95in]{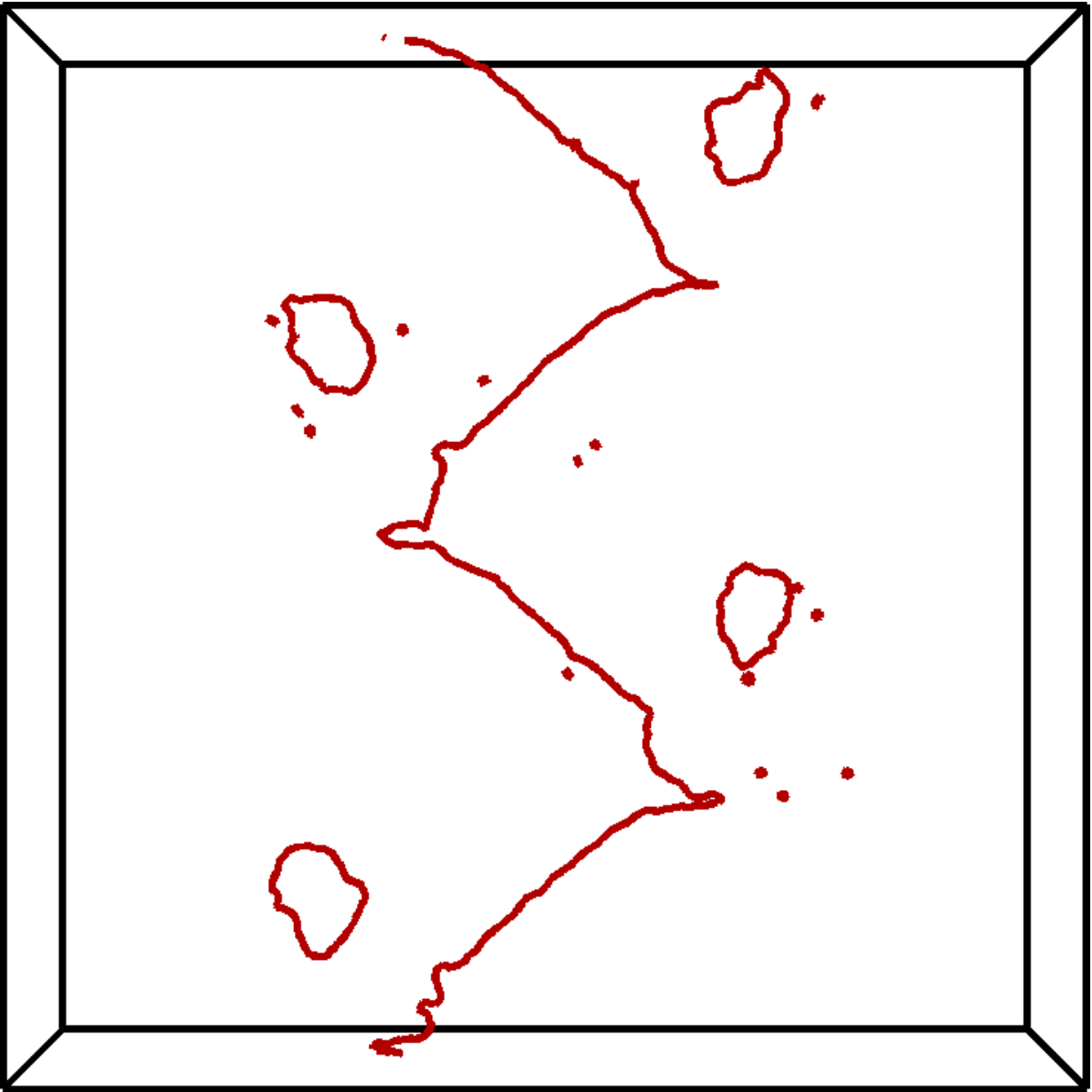}}
  \end{minipage}
 \begin{minipage}[b]{0.49\textwidth}
    \centering
    \subfigure[\label{fig_t2} \hspace{0.3cm}  t = 100 \hspace{1.75cm} t=160 \hspace{1.85cm} t = 240 \hspace{0.8cm}]{
      \label{fig:mini:subfig:a}
      \includegraphics[width=0.95in]{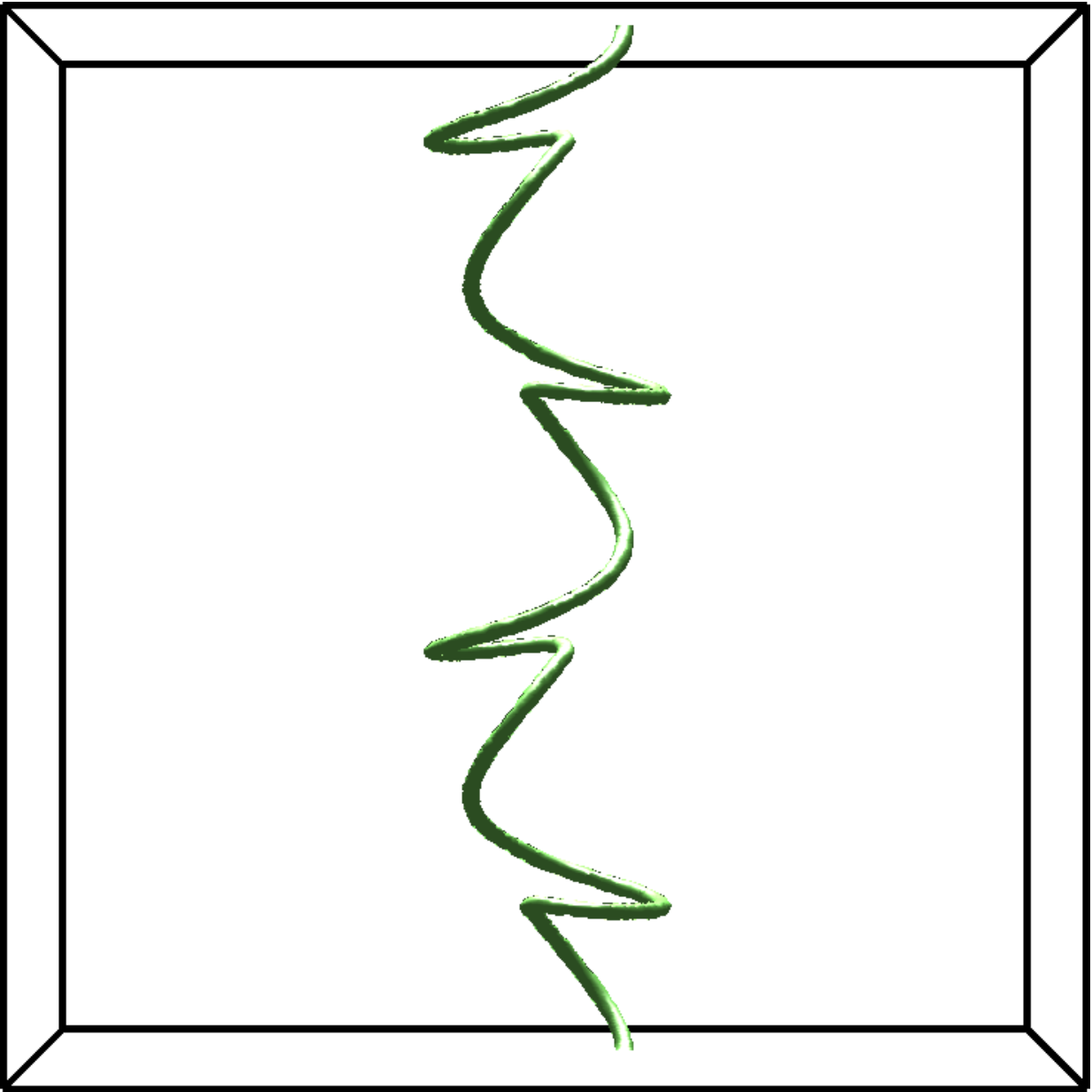}
      \hspace{0.2cm}
      \includegraphics[width=0.95in]{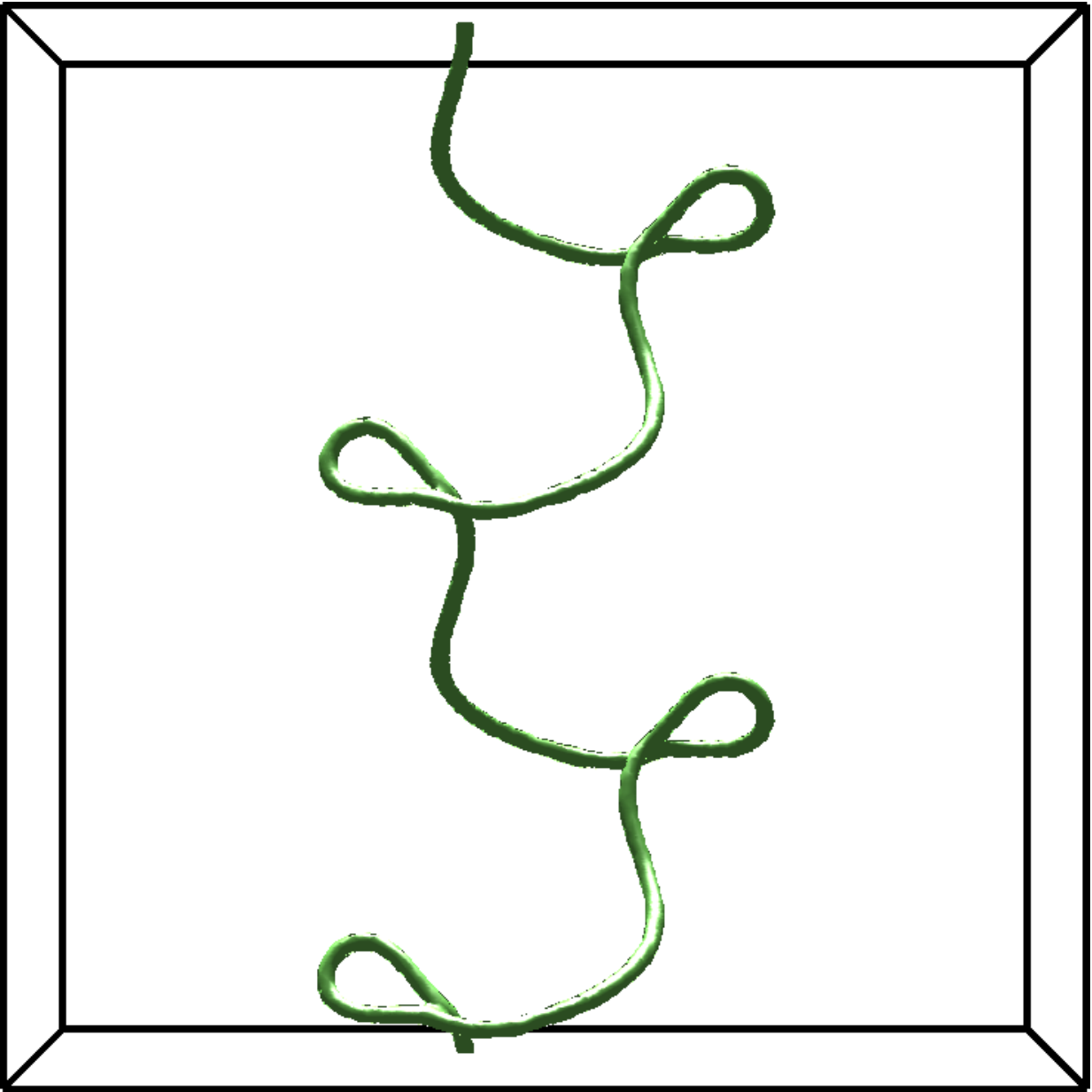}
      \hspace{0.2cm}
      \includegraphics[width=0.95in]{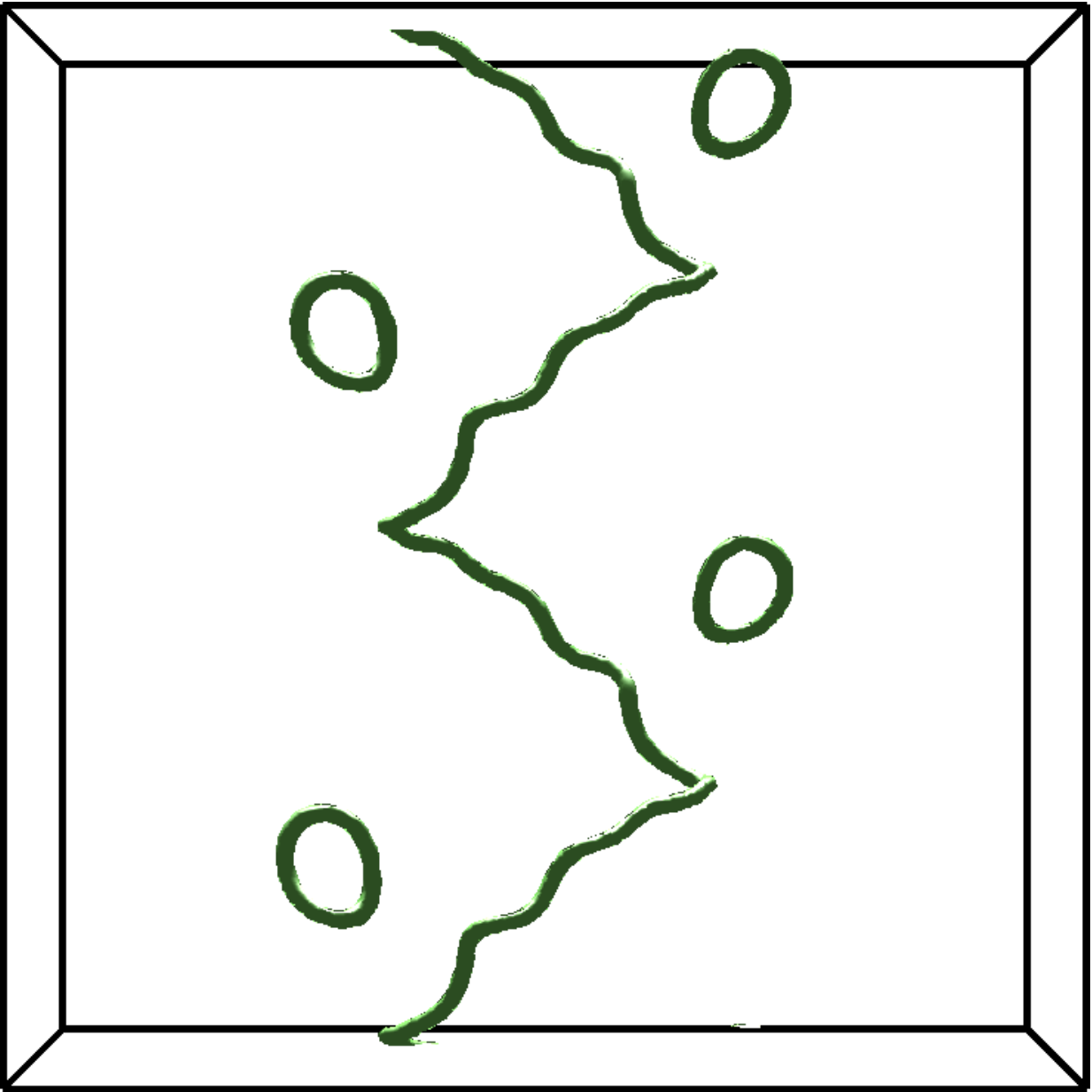}}
  \end{minipage}
 \vspace{-0.3cm}
  \caption{Configuration of vortex filament corresponding to Akhmediev breather at different times computed with (a) LIA (b) Biot-Savart Law (c) Gross-Pitaevskii model. Horizontal and vertical directions correspond to $x$ and $z$ coordinates respectively and depth corresponds to the $y$ coordinate.  \label{fig_BSGP}}
\vspace{-0.4cm}
\end{figure}

An illustration of a breather solution obtained using Eqs.\ (\ref{eqn_Akh})-(\ref{eqn_SF}) is presented in
Fig.\ \ref{fig_BSGP}a which corresponds to $\kappa_o = r/(r^2+h^2)$, and $\tau_o = h/(r^2+h^2)$ where $r = 10$, $h = L/(2\pi n_H)$, $n_H = 6$ and $L=128$ and $\beta = 4\pi$. These set of parameters correspond to a vortex filament which is helical as $t \rightarrow \pm \infty$. The helical vortex can be represented as
\begin{eqnarray*}
x = x_o + r\cos \left( \frac{2\pi n_H s}{AL} \right), \;\;\;\;
y = y_o + r\sin \left( \frac{2\pi n_H s}{AL} \right), \;\;\;\; z = \frac{s}{A},
\end{eqnarray*}
where $A = \left[ 1 + \left( {2\pi n_H}/{L} \right)^2 r^2 \right]^{1/2}$.
Therefore, for the parameters we have chosen, the helical vortex contains 6 waves extending along the $z$-direction over an interval of length $L=128$ and with an amplitude, $r$. This helical vortex approaches the breather solution as $t \rightarrow 0$ and then relaxes back to the helical vortex as $t \rightarrow \infty$. The breather solution is spatially periodic with a wavenumber of the spatial modulation set by $k=4\pi n_B /(L \kappa_o)$ with $n_B=4$ corresponding to the number of breathers that emerge at $t=0$. 
These parameters are, therefore,  closely related to those used in \cite{Umeki10} who also provided an explicit expression for a breather of a vortex filament within the LIA.

In Fig.\ \ref{fig_BSGP}a, we show this solution at $t=-800$, $t=0$ and $t=50$. We note that large amplitude vortex loops form as $t \rightarrow 0$ with the vortex undergoing a self-recrossing. However, since the dynamics are governed by the LIA, the different segments of the vortex pass through one another. Nevertheless, these vortex dynamics suggest that these excitations can act as a key mechanism for creating vortex rings in a more realistic model. This in turn provides a channel
for transferring energy away from a vortex filament. 
One of the advantages of the formulation presented in Eqs.\ (\ref{eqn_Akh})-(\ref{eqn_SF}) is that it allows direct extension of different breather solutions identified for the NLS equation to the case of vortex filaments. 

We will now illustrate that these results apply to vortex dynamics
governed by the full Biot-Savart law. For this purpose, we initialized a vortex
filament at $t=-(3200\pi)$ with identical parameters to the ones considered above. 
For the Biot-Savart simulations, we
set $\Gamma = (4\pi)^2$ and $a_o = 0.91$. By numerically integrating the Biot-Savart law forward in time using the model described in \cite{Baggely11}, we observe in Fig.\ \ref{fig_BSGP}a that the vortex filament recovers the breather type solutions obtained within the LIA. Aside from the difference in speed of propagation of the breathing mode that is seen between the two models, we note that the loops that form on the vortex subsequently lead to a self-reconnection of the vortex and, henceforth, the emission of vortex rings. 

The vortex reconnections in the vortex filament model are enforced during the
computations. In order to model the vortex evolution, including the effect of
reconnections, in a self-consistent way, we illustrate that these
solutions persist even in the mean-field models of a superfluid based
on the Gross-Pitaevskii (GP) equation. We consider the GP equation written in
non-dimensional form as 
\begin{eqnarray}
i\psi_t = -\nabla^2 \psi + |\psi|^2 \psi - \psi
\label{eqn_GP}
\end{eqnarray}
where we have rescaled space and time such that $x \rightarrow x\hbar/\sqrt{2m\mu}$ and $t \rightarrow t \hbar/\mu$,  and $\mu$ is the chemical potential.  In Eq.\ (\ref{eqn_GP}), $\psi(\mathbf{x},t)$ denotes a non-dimensional macroscopic wavefunction of the condensate with the normalisation condition set to $\int |\psi|^2 d^3\mathbf{x}=V$ where $V$ is the volume of the system. We performed simulations in a domain of extent $L_x=L_y= L_z=128$ in the $(x,y,z)$ coordinate directions, respectively. A grid of $129^2 \times 128$ points was used with periodic boundary conditions applied along the $z$-direction and reflecting boundary conditions in the $x$ and $y$-directions. Our vortex, aligned in the $z-$direction, was initialized as in Fig.\ \ref{fig_BSGP}a. This required specifying the location of the zeros of the wavefunction from the breather solutions given above at $t=-3200\pi$. The initial condition for the phase of $\psi$ was then reconstructed by performing a path integral on the velocity field computed from the Biot-Savart law for a corresponding vortex filament. Neumann (reflective) boundary condition were prescribed in the $x$ and $y$-directions with periodic boundary conditions along the $z$-direction.
By relaxing this initial condition through an imaginary time evolution of the GP equation over a short time interval, the phase winding produced a defect in the density field of the condensate thus generating a topological defect in the form of a quantized vortex with the desired configuration. Simulations performed with the GP model and corresponding to the case considered in Fig.\ \ref{fig_BSGP}b are presented in Fig.\ \ref{fig_BSGP}c. The results clearly show
that the breather solutions persist even for these simulations that are based on a mean field description of a superfluid. Indeed, the motion of the quantized superfluid vortices closely follow the evolution obtained from
the vortex filament model. Once the instability enters the nonlinear stages of evolution, the vortex lead to self-reconnections which are correctly modeled by the GP equation since it correctly resolves the inner core of the vortex.

\begin{figure}[t]
\centering
 \begin{minipage}[b]{0.49\textwidth}
    \centering
      \includegraphics[width=0.95in]{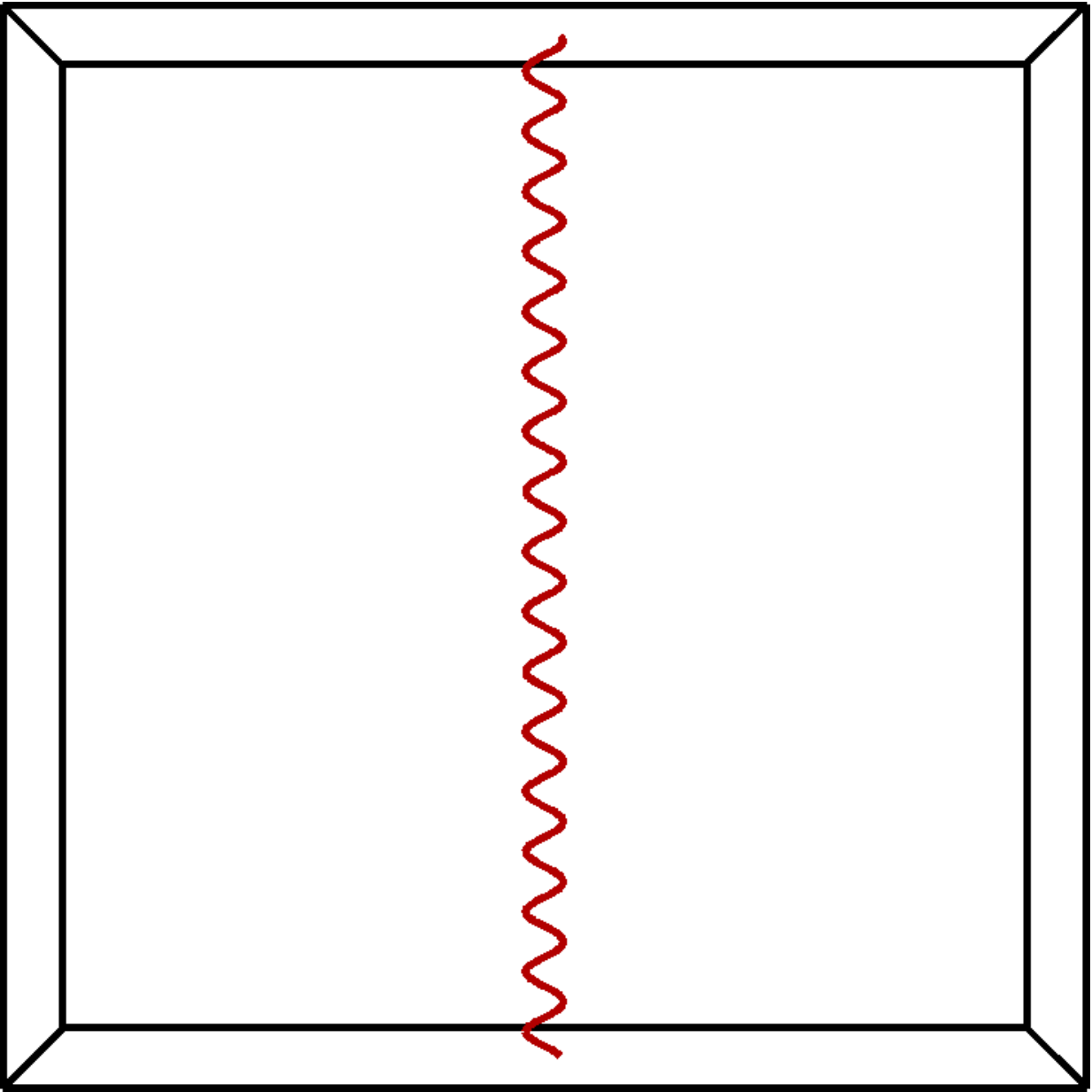}
      \hspace{0.2cm}
      \includegraphics[width=0.95in]{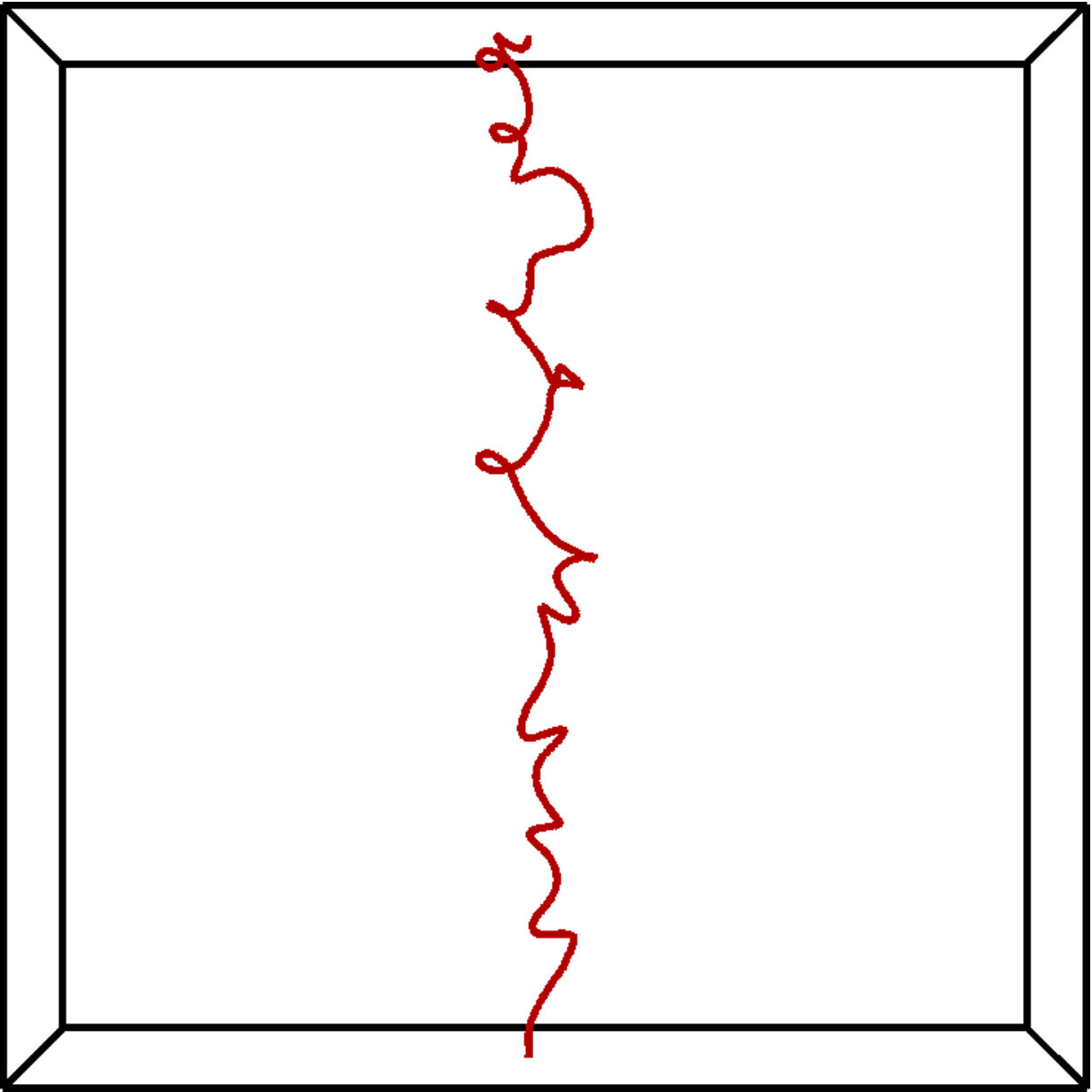}
      \hspace{0.2cm}
      \includegraphics[width=0.95in]{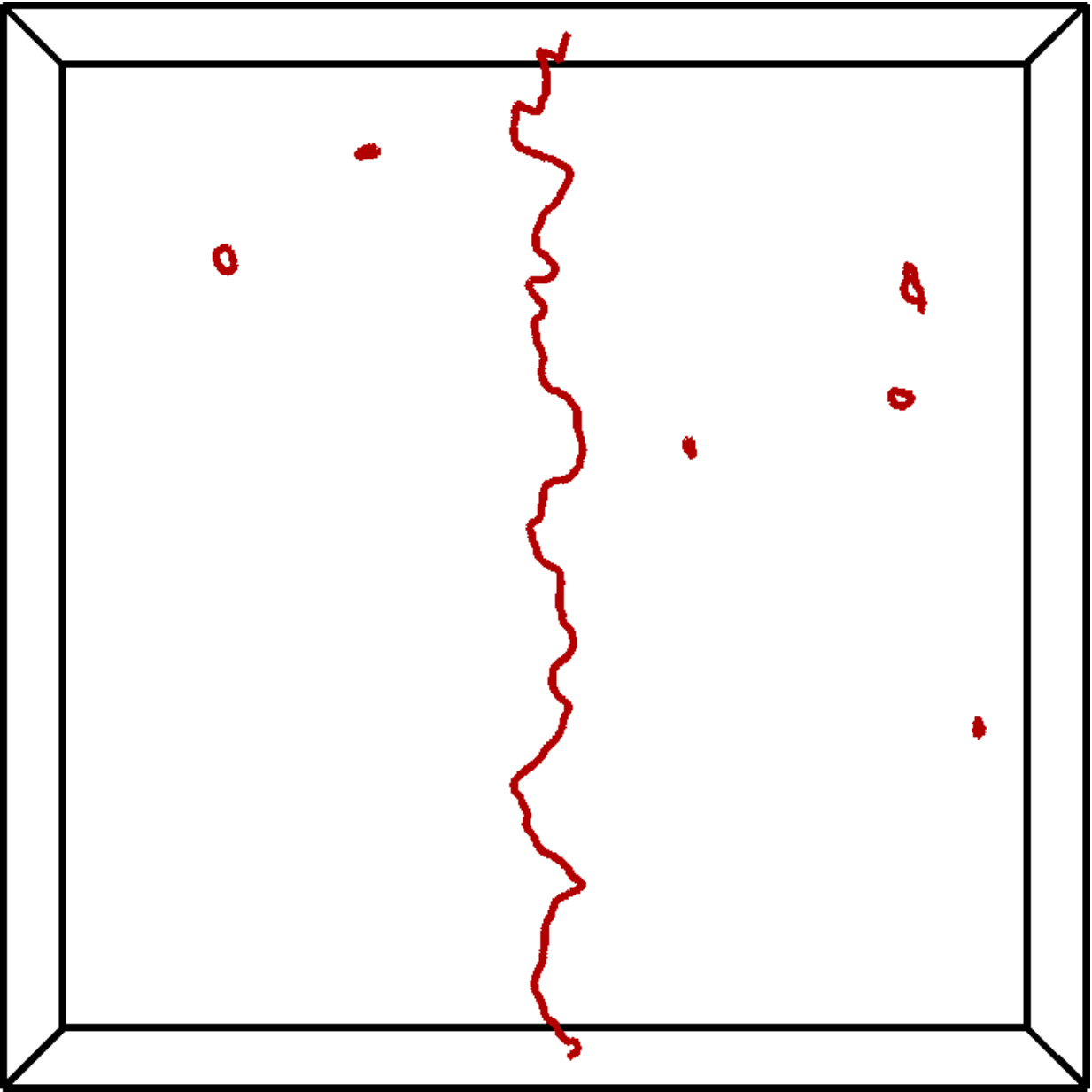} 
      \hspace{0.3cm}
\hspace{0.7cm}  t = 0 \hspace{1.85cm} t= 2650 \hspace{1.75cm} t= 5300 \hspace{0.1cm}      
  \end{minipage}
 \vspace{-0.3cm}
  \caption{Configuration of vortex filament at different times computed with Biot-Savart law for $L=1088$, $n_B=12$, $n_H=17$, and $r=20$. Horizontal and vertical directions correspond to $x$ and $z$ coordinates respectively and depth corresponds to the $y$ coordinate. \label{fig_BS}}
\vspace{-0.4cm}
\end{figure}

The results we have obtained on breathers are expected to be very relevant to our understanding of superfluid turbulence in the ultra-low temperature regime where no normal fluid component is present. Under these conditions, it has been argued in \cite{Svist95,Kozik09} that this scenario can be particularly important in describing superfluid turbulence on a range of length scales where
the large scale coherent motion of polarized vortices, that is responsible for the classical like Kolomogorov cascade \cite{Vinen00,Walmsley08,Alamri08} and which is produced by the large scale forcing, ceases to be relevant. This, so-called, crossover range of scales, in which the quantized nature of superfluid vortices becomes important, has been the subject of debate in terms of how energy cascades to ever increasing wavenumbers \cite{Kozik04,Lvov07} . Given the importance of this mechanism to superfluid turbulence, we have considered a second example of a breather solution with a longer vortex. This allows more unstable  modes to be captured in the simulations which in turn can lead to more
chaotic vortex dynamics that can be important to the self-reconnection scenario in
superfluid turbulence. In Fig.\ \ref{fig_BS}, we present results of a Biot-Savart simulation
corresponding to a breather solution with $n_B=12$ and $n_H=17$, $L=1088$, and $r=20$.
The amplitude of the excitations considered here with $r/2h \sim 0.98$ is consistent with the scenario depicted in \cite{Kozik09} where deposition of energy occurs on scales larger than the self-reconnection driven regime. The resulting vortex dynamics seen in the figure lead to a highly chaotic vortex configuration which ultimately leads to proliferation of vortex rings. We emphasize that in contrast to earlier work where the self-reconnections arise from forcing applied to a vortex filament, our results indicate that the vortex is inherently unstable to the long-wavelength modulational instability which is sufficient to drive reconnections. This is a key observation since the extent of the self-reconnection driven cascade is dependent on the parameter $\Lambda = \ln(R/a_o) \le 15$ for realistic values of $^4$He. The mechanism we have identified suggests that, even for such relatively small values of the parameter, the modulational instability we have identified reinforces the scenario of self-recrossings on these scales. 
Therefore, the generation of vortex loops appears to be an essential mechanism for releasing the excess energy stored in the vortices within the crossover range of scales. Once formed, these vortex loops or rings tend to persist even upon interacting with other vortex lines \cite{Tsubota00} and so can self-propagate towards the boundaries of the containers where they are absorbed and hence energy is dissipated. The remaining energy can then cascade in the form of Kelvin waves. 

In addition to the relevance to superfluid turbulence that we have outlined, perhaps more significant is the fact that these breather solutions persist in field models of topological defects. This suggests that these type of vortex excitations are expected to be relevant to a number of other systems such as cosmic strings and vortons governed by the nonlinear Klein-Gordon equation \cite{Pismen99}. The results we have presented are, therefore, of generic importance to our understanding of the dynamics of topological defects in a broad class of physical systems. Another connection arises for the Heisenberg equation of a 1D ferromagnet. In this case, the equation of motion of the spin vector coincides with the equation of motion for the tangent vector of the vortex filament. Furthermore, we point out that the recent experimental techniques developed by \cite{Kleckner13} on vortex knots may provide a promising means of reproducing the vortex excitations that we have identified under controlled conditions in the laboratory.\\

The author would like to acknowledge Dr.\ A.W.~Baggely for the many helpful discussions regarding
the vortex filament model used in this work.


\begin{thebibliography}{}

\bibitem{Slunyaev11} A.~Slunyaev, I.~Didenkulova, E.~Pelinovsky. Contemp.\ Phys, {\bf 52} 571-590 (2011).

\bibitem{Brand02} J.~Brand, W.P.~Reinhardt, Phys.\ Rev.\ A, {\bf 65}, 043612 (2002).

\bibitem{Ablowitz11} M.~Albowitz. {\em Nonlinear Dispersive Waves: Asymptotic Analysis and Solitons}, Cambridge University Press, (2011).

\bibitem{Jones82} C.A.~Jones, P.H.~Roberts, J.\ Phys.\ A, {\bf 15}, 2599 (1982).

\bibitem{Kuznetsov77} E.~Kuznetsov. Sov.\ Phys.\ Dokl., {\bf 22}, 507 (1977).

\bibitem{Ma79} Y.~Ma. Stud.\ Appl.\ Math, {\bf 60}, 43 (1979).

\bibitem{Akhmediev87} N.N.~Akhmediev, V.M.~Eleonskii, N.E.~Kulagin. Translation
from Teoreticheskaya i Matematicheskaya Fizika, {\bf 72}, 183-196 (1987).

\bibitem{Peregrine83} D.H.~Peregrine, J.\ Austral.\ Math.\ Soc.\ Ser.\ B, {\bf 25}, 16-43 (1983).

\bibitem{Kibler10} B.~Kibler, J.~Fatome, C.~Finot, G.~Millot, F.~Dias, G.~Genty, N.~Akhmediev, J.M.~Dudley. Nature Physics, {\bf 6}(10) 790 (2010).

\bibitem{Chabchoub11} A.~Chabchoub, N.P.~Hoffmann, N.~Akhmediev. Phys.\ Rev.\ Lett., {\bf 106}, 204502 (2011).

\bibitem{Kibler12} B.~Kibler, J.~Fatome, C.~Finot, G.~Millot, G.~Genty, B.~Wetzel, N.~Akhmediev, F.~Dias, J.M.~Dudley. Scientific Reports, {\bf 2} 463 (2012).

\bibitem{Onorato11} M.~Onorato, D.~Proment, A.~Toffoli. Phys.\ Rev.\ Lett., {\bf 107}, 184502 (2011).

\bibitem{Hasimoto72} H.~Hasimoto. J.\ Fluid Mech., {\bf 51}, 477-485 (1972).

\bibitem{Hopfinger82} E.J.~Hopfinger, F.K.~Browand. Nature, {\bf 295}, 393, (1982). 

\bibitem{Cieslinski92} J.~Cie\'sli\'nski. Physics Letters A, {\bf 171}, 323-326 (1992). 

\bibitem{Levi83} D.~Levi, A.~Sym, S.~Wojciechowski. Physics Letters,
{\bf 94A}, 408-411 (1983).

\bibitem{Fukumoto86} Y.~Fukumoto, T.~Miyazaki. J.\ Phys.\ Soc.\ Japan, {\bf 55}, 4152-4155 (1986).

\bibitem{Konno92} K.~Konno, Y.H.~Ichikawa. Chaos, Solitons \& Fractals, {\bf 2}, 237-250 (1992).

\bibitem{Konno91} K.~Konno, M.~Mituhashi, Y.H.~Ichikawa. Chaos, Solitons \& Fractals, {\bf 1}, 55-65 (1991).

\bibitem{Aref84} H.~Aref, M.A.~Stremler. J.\ Fluid Mech., {\bf 148}, 477-497 (1984).

\bibitem{Tsubota09} M.~Tsubota. J.\ Phys.: Condens.\ Matter. {\bf 21}, 164207, (2009).

\bibitem{Vinen08} W.F.~Vinen. Phil.\ Trans.\ R.\ Soc.\ A, {\bf 366}, 2925, (2008).

\bibitem{Donnelly91} R.J.~Donnelly. {\em Quantized Vortices in Helium II}, Cambridge Univ.\ Press, (1991).

\bibitem{Pismen99} L.M.~Pismen. {\em Vortices in Nonlinear Fields: From Liquid Crystals to Superfluids, 
from Nonequilibrium Patterns to Cosmic Strings}, Oxford University Press, (1999).

\bibitem{Svist95} B.V.~Svistunov. Phys.\ Rev.\ B, {\bf 52}, 3647, (1995).

\bibitem{Dysthe99} K.B.~Dysthe, K.~Trulsen. Physica Scripta, {\bf T82}, 48-52 (1999).

\bibitem{Akhmediev86} N.N.~N.N.~Akhmediev, V.I.~Korneev. Translation
from Teoreticheskaya i Matematicheskaya Fizika, {\bf 69}, 189-194 (1986).

\bibitem{Ablowitz90} M.J.~Ablowitz, B.M.~Herbst. SIAM J.\ Appl.\ Math., {\bf 50}, 339-351 (1990).

\bibitem{Benjamin67a} T.B.~Benjamin, J.E.~Feir. J.\ Fluid.\ Mech., {\bf 27},
417-430, (1967).

\bibitem{Benjamin67b} T.B.~Benjamin, K.~Hasselmann. Proc.\ R.\
Soc.\ Lon.\ A, {\bf 299}, 59-76, (1967).

\bibitem{Samuels90} D.C.~Samuels, R.J.~Donnelly. Phys.\ Rev.\ Lett., {\bf 64}, 1385
(1990).

\bibitem{Zakharov68} V.~Zakharov. J.\ Appl.\ Mech.\ Tech.\ Phys., {\bf 9}, 190 (1968).

\bibitem{Umeki10} M.~Umeki. Theor. Comp.\ Fluid Dyn., {\bf 24}, 383-387, (2010).

\bibitem{Baggely11} A.W.~Baggaley, C.F.~Barenghi. Phys.\ Rev.\ B, {\bf 83}, 134509, (2011). 

\bibitem{Kozik09} E.V.~Kozik, B.V.~Svistunov, J.\ Low Temp.\ Phys., {\bf 156}, 215, (2009).

\bibitem{Vinen00} W.F.~Vinen, Phys.\ Rev.\ B, {\bf 61}, 1410, (2000).

\bibitem{Walmsley08}  P.M.~Walmsley, A.I.~Golov. Phys.\ Rev.\ Lett., \textbf{100}, 245301, (2008).

\bibitem{Alamri08} S.Z.~Alamri, A.J.~Youd, C.F.~Barenghi. Phys.\ Rev.\ Lett., {\bf 101}, 215302, (2008).

\bibitem{Kozik04}  E.V.~Kozik, B.V.~Svistunov. Phys.\ Rev.\ Lett., \textbf{92}, 035301, (2004).

\bibitem{Lvov07} V.S.~L'vov, S.V.~Nazarenko, O.~Rudenko. Phys.\ Rev.\ B, {\bf 76}, 024520, (2007).

\bibitem{Tsubota00} M.~Tsubota, T.~Araki, S.K.~Nemirovskii. Phys.\ Rev.\ B, {\bf 62}, 11751, (2000).

\bibitem{Kleckner13} D.~Kleckner, W.T.M.~Irvine. Nature Physics, {\bf 9}, 253, (2013).

 
\end{thebibliography}
\end{document}